\documentclass[ALICE,manyauthors]{cernphprep}
\usepackage[comma,square,numbers,sort&compress]{natbib}
\usepackage{hyperref}
\usepackage{lineno}
\usepackage{xspace}
\usepackage{comment}
\usepackage[T1]{fontenc} 
\usepackage{orcidlink}

\begin{document}
%

\newcommand{\pp}           {pp\xspace}
\newcommand{\ppbar}        {\mbox{$\mathrm {p\overline{p}}$}\xspace}
\newcommand{\XeXe}         {\mbox{Xe--Xe}\xspace}
\newcommand{\PbPb}         {\mbox{Pb--Pb}\xspace}
\newcommand{\pA}           {\mbox{pA}\xspace}
\newcommand{\pPb}          {\mbox{p--Pb}\xspace}
\newcommand{\AuAu}         {\mbox{Au--Au}\xspace}
\newcommand{\dAu}          {\mbox{d--Au}\xspace}

\newcommand{\s}            {\ensuremath{\sqrt{s}}\xspace}
\newcommand{\snn}          {\ensuremath{\sqrt{s_{\mathrm{NN}}}}\xspace}
\newcommand{\pt}           {\ensuremath{p_{\rm T}}\xspace}
\newcommand{\meanpt}       {$\langle p_{\mathrm{T}}\rangle$\xspace}
\newcommand{\ycms}         {\ensuremath{y_{\rm CMS}}\xspace}
\newcommand{\ylab}         {\ensuremath{y_{\rm lab}}\xspace}
\newcommand{\etarange}[1]  {\mbox{$\left | \eta \right |~<~#1$}}
\newcommand{\yrange}[1]    {\mbox{$\left | y \right |~<~#1$}}
\newcommand{\dndy}         {\ensuremath{\mathrm{d}N_\mathrm{ch}/\mathrm{d}y}\xspace}
\newcommand{\dndeta}       {\ensuremath{\mathrm{d}N_\mathrm{ch}/\mathrm{d}\eta}\xspace}
\newcommand{\avdndeta}     {\ensuremath{\langle\dndeta\rangle}\xspace}
\newcommand{\dNdy}         {\ensuremath{\mathrm{d}N_\mathrm{ch}/\mathrm{d}y}\xspace}
\newcommand{\Npart}        {\ensuremath{N_\mathrm{part}}\xspace}
\newcommand{\Ncoll}        {\ensuremath{N_\mathrm{coll}}\xspace}
\newcommand{\dEdx}         {\ensuremath{\textrm{d}E/\textrm{d}x}\xspace}
\newcommand{\RpPb}         {\ensuremath{R_{\rm pPb}}\xspace}

\newcommand{\nineH}        {$\sqrt{s}~=~0.9$~Te\kern-.1emV\xspace}
\newcommand{\seven}        {$\sqrt{s}~=~7$~Te\kern-.1emV\xspace}
\newcommand{\twoH}         {$\sqrt{s}~=~0.2$~Te\kern-.1emV\xspace}
\newcommand{\twosevensix}  {$\sqrt{s}~=~2.76$~Te\kern-.1emV\xspace}
\newcommand{\five}         {$\sqrt{s}~=~5.02$~Te\kern-.1emV\xspace}
\newcommand{\twosevensixnn}{$\sqrt{s_{\mathrm{NN}}}~=~2.76$~Te\kern-.1emV\xspace}
\newcommand{\fivenn}       {$\sqrt{s_{\mathrm{NN}}}~=~5.02$~Te\kern-.1emV\xspace}
\newcommand{\LT}           {L{\'e}vy-Tsallis\xspace}
\newcommand{\GeVc}         {Ge\kern-.1emV/$c$\xspace}
\newcommand{\MeVc}         {Me\kern-.1emV/$c$\xspace}
\newcommand{\TeV}          {Te\kern-.1emV\xspace}
\newcommand{\GeV}          {Ge\kern-.1emV\xspace}
\newcommand{\MeV}          {Me\kern-.1emV\xspace}
\newcommand{\GeVmass}      {Ge\kern-.2emV/$c^2$\xspace}
\newcommand{\MeVmass}      {Me\kern-.2emV/$c^2$\xspace}
\newcommand{\lumi}         {\ensuremath{\mathcal{L}}\xspace}

\newcommand{\ITS}          {\rm{ITS}\xspace}
\newcommand{\TOF}          {\rm{TOF}\xspace}
\newcommand{\ZDC}          {\rm{ZDC}\xspace}
\newcommand{\ZDCs}         {\rm{ZDCs}\xspace}
\newcommand{\ZNA}          {\rm{ZNA}\xspace}
\newcommand{\ZNC}          {\rm{ZNC}\xspace}
\newcommand{\SPD}          {\rm{SPD}\xspace}
\newcommand{\SDD}          {\rm{SDD}\xspace}
\newcommand{\SSD}          {\rm{SSD}\xspace}
\newcommand{\TPC}          {\rm{TPC}\xspace}
\newcommand{\TRD}          {\rm{TRD}\xspace}
\newcommand{\VZERO}        {\rm{V0}\xspace}
\newcommand{\VZEROA}       {\rm{V0A}\xspace}
\newcommand{\VZEROC}       {\rm{V0C}\xspace}
\newcommand{\Vdecay} 	   {\ensuremath{V^{0}}\xspace}

\newcommand{\ee}           {\ensuremath{e^{+}e^{-}}} 
\newcommand{\pip}          {\ensuremath{\pi^{+}}\xspace}
\newcommand{\pim}          {\ensuremath{\pi^{-}}\xspace}
\newcommand{\pipm}          {\ensuremath{\pi^{\pm}}\xspace}
\newcommand{\kap}          {\ensuremath{\rm{K}^{+}}\xspace}
\newcommand{\kam}          {\ensuremath{\rm{K}^{-}}\xspace}
\newcommand{\pbar}         {\ensuremath{\rm\overline{p}}\xspace}
\newcommand{\kzero}        {\ensuremath{{\rm K}^{0}_{\rm{S}}}\xspace}
\newcommand{\lmb}          {\ensuremath{\Lambda}\xspace}
\newcommand{\almb}         {\ensuremath{\overline{\Lambda}}\xspace}
\newcommand{\Om}           {\ensuremath{\Omega^-}\xspace}
\newcommand{\Mo}           {\ensuremath{\overline{\Omega}^+}\xspace}
\newcommand{\X}            {\ensuremath{\Xi^-}\xspace}
\newcommand{\Ix}           {\ensuremath{\overline{\Xi}^+}\xspace}
\newcommand{\Xis}          {\ensuremath{\Xi^{\pm}}\xspace}
\newcommand{\Oms}          {\ensuremath{\Omega^{\pm}}\xspace}
\newcommand{\degree}       {\ensuremath{^{\rm o}}\xspace}

\newcommand{\CKS}          {K$^{*}$(892)$^{\pm}$}
\newcommand{\CKSshort}          {K$^{*\pm}$}
\newcommand{\NKSshort}          {K$^{*0}$}
\newcommand{\NKS}          {K$^{*}$(892)$^{0}$}
\newcommand{\KS}{K$\mathrm{^{0}_{S} }$}
\newcommand{\ENfive}        {$\sqrt{\it{s}_{\mathrm{NN}}} =$ 5.02 TeV\xspace}
\newcommand{\ENthirteen}    {$\sqrt{\it{s}} =$ 13 TeV\xspace}
\newcommand{\ENtwo}        {$\sqrt{\it{s}_{\mathrm{NN}}} =$ 2.76 TeV\xspace}
\newcommand{\CKSDC}     {K$^{*\pm} \rightarrow \mathrm{K^{0}_{S} \pi^{\pm}}$\xspace}
\newcommand{\CKSDClong}     {K$^{*}$(892)$^{\pm} \rightarrow \mathrm{K^{0}_{S} \pi^{\pm}}$\xspace}
\newcommand{\NKSDClong}     {K$^{*}$(892)$^{0} \rightarrow \mathrm{K^{\mp} \pi^{\pm}}$\xspace}
\newcommand{\CKSDCKL}     {K$^{*\pm} \rightarrow \mathrm{K^{0} \pi^{\pm}}$\xspace}
\newcommand{\RAA}{$R_{\mathrm{AA}}$\xspace}
\newcommand{\pion}          {$\mathrm{\pi}$}
\newcommand{\fone}          {f$_{1}$(1285)}
\newcommand{\foneshort}          {f$_{1}$}
\newcommand{\foneDC}     {f$_{1} (1285) \rightarrow \mathrm{K^{0}_{S} K^{\pm}\pi^{\mp}}$\xspace}

\begin{titlepage}
\PHyear{2024}       
\PHnumber{239}      
\PHdate{16 September}  

\title{Measurement of \textbf{f}$_{1}$(1285) production in pp collisions at $\sqrt{\textbf{\textit{s}}}$ = 13 TeV}
\ShortTitle{f$_{1}$(1285) resonance production in ALICE}   

\Collaboration{ALICE Collaboration\thanks{See Appendix~\ref{app:collab} for the list of collaboration members}}
\ShortAuthor{ALICE Collaboration} 

\begin{abstract}
	
This study presents the first measurement of the f$_{1}$(1285) resonance using the ALICE detector in inelastic proton--proton collisions at a center-of-mass energy of 13 TeV. The resonance is reconstructed at midrapidity ($|y| <$ 0.5) through the hadronic decay channel f$_{1} (1285) \rightarrow \mathrm{K^{0}_{S} K^{\pm}\pi^{\mp}}$. Key measurements include the determination of its mass, transverse-momentum integrated yield, and average transverse momentum. Additionally, the ratio of the transverse-momentum integrated yield of f$_{1}$(1285) to pion is compared with calculations from the canonical statistical hadronization model. The model calculation, assuming a zero total strangeness content for f$_{1}$(1285), reproduces the data within 1$\sigma$ deviation, shedding light on the quark composition of f$_{1}$(1285).

\end{abstract}
\end{titlepage}

\setcounter{page}{2} 


\section{Introduction} 

Quantum chromodynamics (QCD), the theory that governs the strong force, describes how colored quarks and gluons interact, forming various types of hadronic states. This includes mesons, which consist of quark--antiquark pairs, and baryons, composed of three quarks or antiquarks. Beyond these conventional structures, there is a growing interest in exotic states like tetraquarks and pentaquarks, which feature unconventional quark combinations~\cite{Belle:2003nnu, Belle:2007hrb, BESIII:2013ris, CMS:2013jru, LHCb:2014zfx, LHCb:2019kea, LHCb:2021chn, ALICE:2021ovd, ALICE:2023eyl}. Investigations into exotic states can be traced back to the early development of the constituent quark model, which serves as a fundamental framework for understanding the composition of hadrons~\cite{Gell-Mann:1964ewy, Zweig:1964jf, Jaffe:1976ig, Lipkin:1987sk}.

An exemplary candidate for an exotic particle under consideration is the \fone\xspace meson~\cite{ParticleDataGroup:2022pth}. Aligned within the quark model as a member of the $^{3}P_{1}$ axial-vector nonet, the \fone\xspace was independently discovered in p$\overline{\rm p}$ annihilation experiments at BNL~\cite{Miller:1965zza} and CERN~\cite{dAndlau:1965cww} in 1965. Both experiments observed a resonance decaying to K$\overline{\rm K}\pi$ with quantum numbers $I^{G}(J^{PC})$ = 0$^{+}(1^{++})$. Low-energy experiments have provided essential insights into the production and decay mechanisms of \fone\xspace through various processes, including hadronic decays, photoproduction, and central exclusive production. The \fone\xspace state has been observed in pp collisions by the WA102~\cite{WA102:1997gkz, WA102:1998zhh, WA102:1997sum} and WA76~\cite{WA76:1991kit} experiments at CERN, E690 at Fermilab~\cite{E690:1997tev}, and by the L3 Collaboration with $\gamma\gamma$ collisions at CERN~\cite{L3:2000gjc, L3:2001cyf}. Additionally, it has been observed in hadronic Z decays at LEP~\cite{Gavillet:2001ym}, in photoproduction from a proton target with CLAS data~\cite{CLAS:2016zjy}, and beauty-hadron decays at LHCb~\cite{LHCb:2013ged}. However, despite these extensive observations, the precise quark composition of the \fone\xspace remains elusive. Despite these extensive observations, among which few suggests a predominantly non-strange meson composition~\cite{WA102:1997sum, Gavillet:2001ym}, the precise quark composition of the \fone\xspace remains elusive. Theoretical predictions regarding the valence quark content of the \fone\xspace meson are broadly classified into three categories: (i) as a bound state comprising of light up (u) and down (d) quarks, (ii) as a bound state formed by both light and strange (s) quarks, and (iii) as molecular configurations involving K$\overline{\rm K}^{\rm*}$~\cite{Aceti:2015pma}. Quark composition of the \fone\xspace meson involving only light quarks can be expressed as a linear combination of u and d quarks, $\rm \frac{1}{\sqrt{2}}(\bar{u}u + \bar{d}d)$~\cite{Osipov:2017ray}, whereas the presence of strange quarks in the \fone\xspace meson gives three different possibilities of quark compositions: tetraquark state $\rm \frac{1}{\sqrt{2}}(su\bar{s}\bar{u} + sd\bar{s}\bar{d}$)~\cite{Kanada-Enyo:2005gga}, bound state of light quarks with a mixture of strange quarks ($\rm \frac{\alpha}{\sqrt{2}}(\bar{u}u + \bar{d}d)$ + $\rm \delta s\bar{s}$)~\cite{Close:1997nm}, and the bound state of light quarks with a mixture of strange quarks and gluons ($\rm \frac{\alpha }{\sqrt{2}}(\bar{u}u + \bar{d}d)$ + $\rm \delta_{1} s\bar{s}$ + $\delta_{2} G$)~\cite{Moreira:2017ulo}, where G is the gluon state. Here $\alpha, \delta$, $\delta_{1}$, and $\delta_{2}$ are the Clebsch Gordan Coefficients of appropriate value. Regardless of the specific composition, the net strangeness of the \fone\xspace meson remains zero in all these scenarios. Recently, the LHCb collaboration measured the branching fraction ratio of $\bar{\mathrm{B}}^{0}$$\rightarrow$J/$\psi$\fone\xspace to $\bar{\mathrm{B}}^{0}_{\mathrm{s}}$$\rightarrow$J/$\psi$\fone\xspace, obtaining a value of 11.6 $\pm$ 3.1\%. This result deviates from the tetraquark structure interpretation of the \fone\xspace meson, with a significance of 3.3$\sigma$\cite{LHCb:2013ged}. The underlying quark content of \fone\xspace is expected to influence its yield~\cite{ExHIC:2011say}. Notably, calculations using the canonical-ensemble-based Statistical Model ($\gamma_{\rm S}$CSM)~\cite{Vovchenko:2019kes} reveal significant differences in hadron yields based on their strangeness content~\cite{ALICE:2022qnb}. The study reported in this Letter explores the strangeness content of the \fone\xspace meson by comparing its transverse-momentum (\pt) integrated yield obtained from ALICE data with $\gamma_{\rm S}$CSM calculations.

In high-energy heavy-ion collisions, compelling evidences for the formation of a strongly-interacting quark--gluon plasma (QGP) have been observed~\cite{STAR:2000ekf, STAR:2005gfr, STAR:2003pjh, STAR:2002svs, STAR:2003wqp, PHENIX:2001hpc, PHENIX:2004vcz, BRAHMS:2004adc, PHOBOS:2004zne, ALICE:2022wpn, ALICE:2010suc, ALICE:2010yje, ALICE:2011ab, Heinz:2008tv, Niida:2021wut}. This deconfined and strongly interacting state expands and cools down as a nearly perfect liquid~\cite{Bernhard:2019bmu} until the temperature reaches the pseudo-critical temperature of approximately $155$~MeV~\cite{HotQCD:2018pds}. After this phase, a transition to confined QCD matter occurs which creates a hot and dense gas of interacting hadrons. Within this environment resonances decay and particles interact (pseudo)elastically until they decouple~\cite{ALICE:2023ifn}. At the LHC, the system produced in Pb--Pb collisions undergoes decoupling after approximately $10$ fm/$c$~\cite{ALICE:2011dyt}. The study of hadronic resonances with varying lifetimes is crucial for characterizing the late hadronic stage of the collision. Depending on the lifetime of resonances, rescattering and regeneration processes affect their yield~\cite{ALICE:2021ptz, ALICE:2014jbq, NA49:2008goy, NA49:2000jee, NA49:2011bfu, PHENIX:2014kia, PHENIX:2022rvg, PHENIX:2022hku, PHENIX:2004spo, STAR:2008inc, STAR:2008bgi, STAR:2002npn, STAR:2004bgh, STAR:2010avo}. Given that \fone\xspace has a lifetime of approximately 8.7 fm/$c$~\cite{ParticleDataGroup:2022pth}, placing it between the lifetimes of K$^{*0}$ meson and $\Lambda^{*}$ baryon, it becomes an indispensable component for systematically studying rescattering effects and properties of the hadronic phase in heavy-ion collisions. Furthermore, theoretical studies suggested that the \fone\xspace meson could be pivotal in exploring the partial restoration of chiral symmetry within the nuclear medium~\cite{Gubler:2016djf}. It has been found that the \fone\xspace, a chiral partner of the $\omega$ meson, could exhibit a significant mass shift from its vacuum expectation ($1281.9$ $\pm$ $0.5$ MeV/$c^{2}$) in the presence of finite baryon density. Similar trends for chiral partners are predicted at the high temperatures reached in heavy-ion collisions at LHC energies~\cite{Rapp:1999ej}. Searches for (partial) chiral symmetry restoration effects are typically investigated through the electromagnetic decays of vector mesons, as they are not affected by rescattering, unlike hadronic decays. However, since the \fone\xspace meson is not particularly broad, rescattering effects may be less dominant in this case. Additionally, by performing measurements in peripheral Pb--Pb collisions, such effects could be further minimized. Another important aspect of the measurement is the yield ratio of the \fone\xspace meson to its chiral partner, the $\omega$, which can provide valuable insights into chiral symmetry restoration. This yield ratio is expected to approach unity~\cite{Sung:2023oks} as one moves towards more peripheral Pb--Pb collisions due to the mass degeneracy of the chiral partners.
Therefore, measurements of the \fone\xspace production in pp collisions are crucial to constitute a reference for studying the partial restoration of chiral symmetry and rescattering effects in heavy-ion collisions.

This Letter presents the first measurement of the inclusive production of the \fone\xspace resonance at midrapidity ($|y|<$ 0.5) in inelastic pp collisions at a center-of-mass energy $\sqrt{s}$ of 13 TeV. The article is structured as follows: Section~\ref{EAP} outlines the ALICE experimental setup, Section~\ref{DS} details the event and track selection criteria, Section~\ref{DAT} presents the data analysis technique, and Section~\ref{SU} describes the study of systematic uncertainties. Results are presented in Section~\ref{RE}, and the Letter concludes with a summary in Section~\ref{Summ}.

\section{Experimental apparatus} \label{EAP}

The yield of the \fone\xspace meson is measured in pp collisions at \ENthirteen using data collected by the ALICE detector. A detailed description of the ALICE detector and its performance can be found in Refs.~\cite{ALICE:2008ngc, ALICE:2014sbx}. Several key detectors, including the Inner Tracking System (ITS)~\cite{ALICE:2010tia}, Time Projection Chamber (TPC)~\cite{Alme:2010ke}, Time-of-Flight (TOF)~\cite{ALICE:2000xcm, ALICE:2002imy}, and V0~\cite{ALICE:2013axi} detectors, have been used for the analysis presented in this Letter.

For event triggering and mitigating beam-induced background effects, the V0 detector is used. It consists of two scintillator arrays, V0A and V0C, which are positioned on either side of the interaction point along the beam line and cover the pseudorapidity intervals 2.8 $ < \eta <$ 5.1 and $-$3.7 $ < \eta <$ $-$1.7, respectively. The minimum bias trigger used in this analysis is defined by coincident signals in the V0A and V0C detectors. 

The ITS and TPC detectors, housed within a 0.5 T solenoidal magnet, play crucial roles in tracking and identifying charged particles and reconstructing primary and secondary vertices. The ITS and TPC cover a pseudorapidity range of $|\eta| < $ 0.9 and the full azimuthal angle.

The ITS comprises six cylindrical silicon layers surrounding the beam vacuum tube. The two innermost layers are formed by Silicon Pixel Detectors (SPD), followed by two layers of Silicon Drift Detectors and two layers of Silicon Strip Detectors. The ITS is crucial for determining primary and secondary vertices. Additionally, the ITS improves the momentum and angle resolution for charged particles reconstructed by the TPC.

The TPC serves as the core of the ALICE detector~\cite{Alme:2010ke, ALICE:2014sbx}. It is a large cylindrical drift detector, spanning radial and longitudinal ranges of approximately 85 $<~r~<$~250 cm and $-$250 $<~z~<$~250 cm, respectively. The endcaps of the TPC incorporate multiwire proportional chambers segmented radially into pad rows. The TPC provides three-dimensional
spatial information for up to 159 tracking points. Charged tracks originating from the primary vertex can be reconstructed down to \pt $\sim$ 150 MeV/$c$. The particle identification is based on the specific energy loss (d$E$/d$x$) in the TPC, which is measured with a resolution of 5\% in pp collisions~\cite{Alme:2010ke}. The measured d$E$/d$x$ is compared with the expected value for a given particle species calculated with a Bethe--Bloch parameterization.

The TOF is placed outside the TPC and employs Multigap Resistive Plate Chambers, covering the pseudorapidity range of $|\eta|<$0.9 and full azimuthal angle. The TOF detector identifies particle species at intermediate \pt\ via measurements of their time-of-flight from the interaction point to the TOF detector with a time resolution of 80 ps in pp collisions~\cite{ALICE:2000xcm}.

\section{Data sample, event and track selections} \label{DS}

The data utilized in the present analysis were collected by the ALICE detector in 2016, 2017, and 2018. The position of the primary vertex along the beam axis ($z$-axis of the ALICE reference frame) is required to be within 10 cm from the nominal center ($z=$~0) of the ALICE detector. As detailed in Refs.~\cite{ALICE:2014sbx, ALICE:2019avo}, offline event selections are applied to reduce the beam-induced background and pileup events. After applying the event selection criteria, approximately 1.5 billion minimum-bias events (corresponding to an integrated luminosity of 32.08 $\pm$ 0.51 nb$^{-1}$~\cite{alice2021alice}) have been analyzed for this \fone\xspace measurement.

Given the short-lived nature of the \fone\xspace meson, its reconstruction is performed through the hadronic decay channel, \foneDC, with a branching ratio (BR) of (2.25 $\pm$ 0.1)$\%$~\cite{ParticleDataGroup:2022pth}. The BR value is computed from the one of K$\overline{\rm K}\pi$ reported in~\cite{ParticleDataGroup:2022pth} accounting for all possible combinations of kaons and pions and 50\% probability that K$^{0}$ is \KS\xspace. The analysis is performed in the transverse momentum range of 1 $< p_{\rm T} <$ 12 GeV/$c$ at midrapidity ($|y| <$ 0.5). At lower \pt ($<$ 1 GeV/$c$), the \fone\xspace signal is not statistically significant because of the presence of large backgrounds.

Charged tracks are reconstructed using the ITS~\cite{ALICE:2013nwm} and TPC~\cite{Alme:2010ke} detectors. To ensure high track quality, the standard track selection criteria~\cite{ALICE:2019etb, ALICE:2020jsh} are employed in this work. Charged tracks originating from the primary vertex are required to satisfy \pt$>$ 0.15 GeV/$c$ and $|\eta|<$ 0.8 for uniform acceptance. Selected tracks need to have two hits in the ITS, of which at least one hit in the SPD, and traverse radially a minimum of 70 out of the total 159 pad rows of the TPC. The maximum $\chi^{2}$ per space point in the TPC and ITS, obtained from the track fit, is required to be 4 and 36, respectively. To mitigate the contamination of secondary charged particles, the distance of closest approach in the transverse plane of reconstructed tracks to the primary vertex (DCA$_{xy}$) is required to be smaller than 7$\sigma_{\mathrm{DCA}_{xy}}$, where $\sigma_{\mathrm{DCA}_{xy}}$ denotes the DCA$_{xy}$ resolution. The \pt-dependent DCA$_{xy}$ resolution is parameterized as $\sigma_{\mathrm{DCA}_{xy}} = 0.0105+0.0350/(\pt\mathrm{/(GeV/}c))^{1.1}$ cm~\cite{ALICE:2020jsh}. The DCA to primary vertex in the longitudinal direction is constrained to be within 2 cm. The detected charged particles are identified using information from the TPC and TOF detectors~\cite{ALICE:2000xcm}. In the TPC, particle identification is based on their specific ionization energy loss (d$E$/d$x$), ensuring that pions and kaons have a specific energy loss within 2 standard deviations ($\sigma_{\mathrm{TPC}}$) from the expected d$E$/d$x$ values derived from the Bethe--Bloch parameterization. Here, $\sigma_{\mathrm{TPC}}$ represents the TPC’s d$E$/d$x$ resolution~\cite{Alme:2010ke}. In the TOF, identification relies on the measured time of flight, which must be within 3$\sigma_{\mathrm{TOF}}$ of the expected value for each particle species, provided the track has a hit in the TOF~\cite{ALICE:2002imy}. If a track lacks a hit in the TOF, identification is carried out using only the TPC.

The \KS\xspace is reconstructed through its weak decay topology (V$^{0}$ topology)~\cite{ALICE:2021xyh}, via the \KS\xspace $ \rightarrow \mathrm{\pi^{-} \pi^{+}}$ decay channel with a BR of (69.2 $\pm$ 0.05)\%~\cite{ParticleDataGroup:2022pth}. The selection criteria for \KS\xspace reconstruction are detailed in Table~\ref{tab:K0ssel}. Two oppositely-charged pions produced from the \KS\xspace decay are identified with the 4$\sigma_{\mathrm{TPC}}$ requirement in the acceptance window $|\eta|<$ 0.8. The distance of closest approach between negatively and positively charged tracks (DCA$_{\pi^{-}\pi^{+}}$) is required to be less than 1.0 cm. Additionally, the DCA of charged tracks and V$^{0}$ to the primary vertex must be greater than 0.06 cm and less than 0.3 cm, respectively. The cosine of the pointing angle, representing the angle between the V$^{0}$ momentum and the line connecting the secondary to the primary vertex, has to be greater than 0.97. Only \KS\xspace candidates whose secondary vertex radial position is larger than 0.5 cm are selected to reconstruct \fone. Furthermore, candidates with a proper lifetime $LM_{\rm{K^{0}_{S}}}/p$ greater than 15 cm/$c$ are excluded. Here, $L$ represents the linear distance between the primary and secondary vertex, $M_{\rm{K^{0}_{S}}}$ is the world-average mass~\cite{ParticleDataGroup:2022pth} of \KS, and $p$ indicates the total momentum of \KS. An additional selection, called "Competing V0 rejection" or veto on $\Lambda$ invariant mass, is applied by recalculating the V0 mass, assuming that one of two pions is a proton. If the recalculated mass is compatible with the $\Lambda$ mass within 4.3 MeV/$c^{2}$, which is about three times the width of the $\Lambda$ invariant mass peak in ALICE~\cite{ALICE:2013wgn, ALICE:2021xyh, ALICE:2020jsh}, the selected particle is rejected. Finally, the invariant mass of $\mathrm{\pi^{+}\pi^{-}}$ must be compatible within 6$\sigma_{\rm M}$ of the \KS\xspace nominal mass, where $\sigma_{M}$ is the width of the \KS\xspace invariant mass peak and is about 5~MeV/$c^{2}$. The \KS\xspace candidates that satisfy the aforementioned topological selection criteria at midrapidity ($|y| <$ 0.5) are used in the reconstruction of the \fone\xspace resonance.

\begin{table}[ht]
	\centering
	\caption{Selection criteria for \KS.}
	\begin{tabular}{ll}	
		\hline
		Selection criteria & Value\\
		\hline
		TPC crossed rows & $>$ 70\\
		Acceptance window of pions ($|\eta|$) & $<$ 0.8\\
		n$\sigma_{\mathrm{TPC}}$ for $\pi^{\pm}$ & $<$ 4\\
		DCA$_{\pi^{-}\pi^{+}}$ & $<$ 1.0 cm\\
		DCA of V$^{0}$ daughters to PV & $>$ 0.06 cm\\
  		DCA of V$^{0}$ to PV & $<$ 0.3 cm\\
		V$^{0}$ cosine pointing angle & $>$ 0.97\\
		V$^{0}$ radius & $>$ 0.5 cm\\
		Proper lifetime & $<$ 15 cm/$c$\\
		Veto on $\Lambda$ invariant mass & $>$ 4.3 MeV/$c^{2}$\\
		\KS\xspace mass window (in units of $\sigma_{M}$) & $\pm$ 6\\
        \hline
	\end{tabular}
	\label{tab:K0ssel}	
	
\end{table}

\section{Data analysis} \label{DAT}

\begin{figure}[!hbt]
	\centering
	\begin{minipage}{.5\textwidth}
		\includegraphics[height=1.0\linewidth,width=1.0\linewidth]{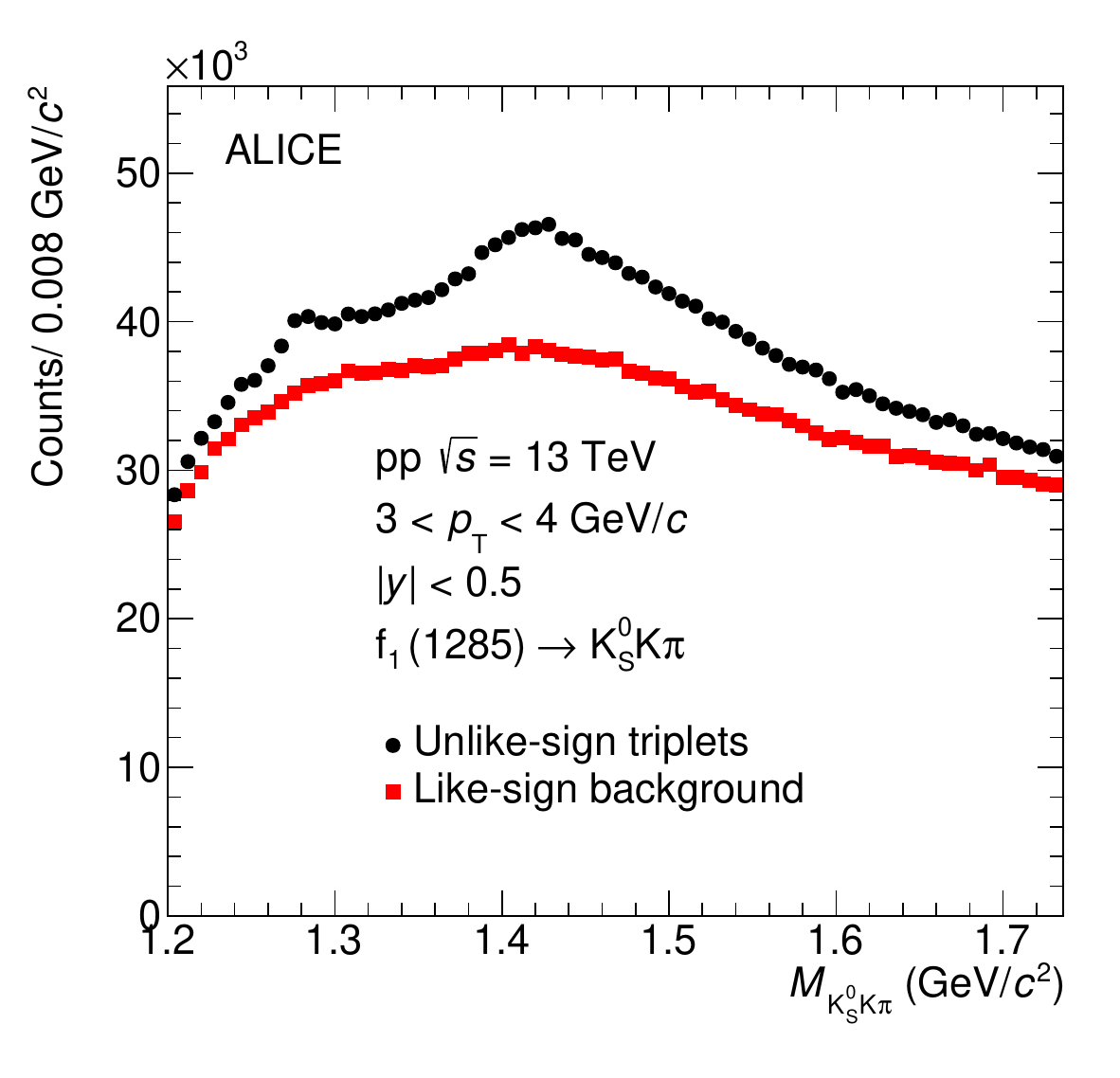}
	\end{minipage}%
	\begin{minipage}{.5\textwidth}
		\centering
		\includegraphics[height=1.0\linewidth,width=1.0\linewidth]{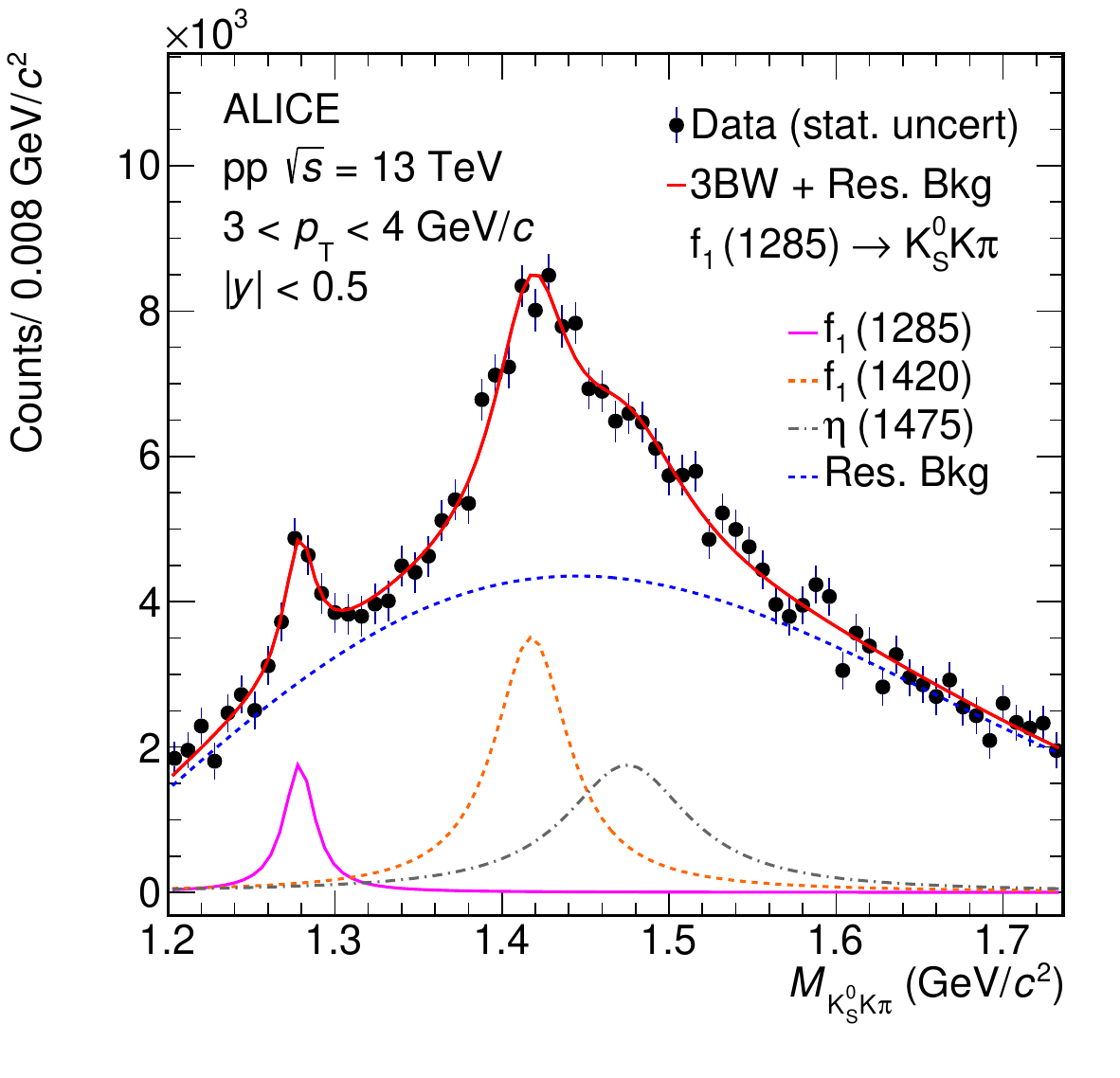}
	\end{minipage}	
	\caption{\label{fig:invmass} Like- and unlike-sign (left) and the like-sign-subtracted (right) invariant mass
distribution of \KS K$\pi$ triplets in $|y|<$ 0.5 in minimum-bias pp collisions at \s = 13 TeV. The subtracted distribution is fitted with the function defined by Eq.~\ref{eq:Fitfunction}, and the dotted blue line describes the residual background distribution, which is given by Eq.~\ref{eq:Resbkg}.}
\end{figure}

The reconstructed \KS\xspace are paired with charged kaons forming a \KS K$^{\pm}$ pair. This \KS K$^{\pm}$ pairs are combined with oppositely charged pions to reconstruct the \fone\xspace resonance. To enhance the significance of the \fone\xspace signal, the invariant mass of the \KS K$^{\pm}$ pair is required to be below 1040 MeV/$c^{2}$. The invariant-mass distribution of \KS K$^{\pm} \pi^{\mp}$ triplets accommodates all resonances that decay into \KS K$^{\pm} \pi^{\mp}$ as well as substantial combinatorial background, as can be seen in the invariant mass distribution of unlike-sign combinations in the f$_{1}$-candidate \pt interval 3 $<$ \pt $<$ 4 GeV/$c$, shown by the black markers in the left panel of Fig.~\ref{fig:invmass}. The combinatorial background is estimated using like-sign \KS K$^{\pm} \pi^{\pm}$ triplets~\cite{STAR:2004bgh, ALICE:2019etb} (red markers in left panel of Fig.~\ref{fig:invmass}). The right panel of Fig.~\ref{fig:invmass} presents the invariant mass distribution of the like-sign-subtracted \KS K$^{\pm} \pi^{\mp}$ triplets for 3 $<$ \pt $<$ 4 GeV/$c$ in pp collisions at \s = 13 TeV. After the subtraction, three resonances, i.e, \fone, f$_{1}$(1420), and $\eta$(1475) can be identified in the considered invariant mass range along with a residual background of correlated pairs. Theoretical models based on K$^{*}\overline{\rm K}$ dynamics~\cite{Aceti:2015pma, Debastiani:2016xgg} offer intriguing insights into the nature of f$_{1}$(1420), which could be explored in future studies.

The correlated background mainly arises from jets and decays of resonances with misidentification and multiple decay chains~\cite{STAR:2004bgh}.
The like-sign-subtracted invariant-mass distribution is fitted assuming a sum of three non-relativistic Breit--Wigner distributions~\cite{Gavillet:2001ym, STAR:2004bgh} for the \fone, f$_{1}$(1420), and $\eta$(1475) mesons and an additional residual background~\cite{Gavillet:2001ym}. The fit function is given by
\begin{equation}
	\frac{\mathrm{d}N}{\mathrm{d}M} = \sum_{i=1}^{3}\frac{Y_{i}}{2\mathrm{\pi}} \frac{\Gamma_{i}}{(M - M_{i})^{2} + \Gamma^{2}_{i}/4} + \mathrm{f}_\mathrm{Res. Bkg} (M),
	\label{eq:Fitfunction}
\end{equation}

where the index $i$ ranges over \fone, f$_{1}$(1420), and $\eta$(1475) resonances. The $M_{i}$, $\Gamma_{i}$, and $Y_{i}$ parameters denote the masses, widths, and normalization constants of these three resonances, respectively. The $M$ corresponds to the invariant mass of the \KS K$^{\pm} \pi^{\mp}$ ($M_{\rm{K^{0}_{S} K \pi}}$) triplets. The mass resolution of the detector for the reconstruction of \fone\xspace is negligible as compared to its vacuum width (22 $\pm$ 1) MeV/$c^{2}$~\cite{ParticleDataGroup:2022pth} and is not included in the fit function. To ensure the robustness of the fit, an alternative modeling using a relativistic Breit-Wigner parameterization was performed, yielding consistent results within uncertainties. The residual background function~\cite{Gavillet:2001ym} is expressed as\\
\begin{equation}
	 \mathrm{f}_\mathrm{Res. Bkg} (M) = [M- (m_{\mathrm{\pi}}+M_{\mathrm{K_{S}^{0} K}})]^{n} \exp(-AM - BM^{2}),
	\label{eq:Resbkg}
\end{equation}
where $m_{\mathrm{\pi}}$ is the mass of $\pi$ meson and $M_{\mathrm{K_{S}^{0}K}}$ is the invariant mass of the \KS K pair. Here, $A$, $B$, and $n$ are the fit parameters. The width parameters of the Breit--Wigner functions are fixed to their world-average values~\cite{ParticleDataGroup:2022pth} in the standard fit case, which are 22, 54, and 90 MeV/$c^{2}$, respectively. The masses and the normalization constants of the three resonances are left free. Finally, the raw yields of \fone\xspace in each \pt interval are obtained from the integral of the Breit--Wigner distribution, as done in Refs.~\cite{ALICE:2019xyr, ALICE:2019etb}. 

\begin{figure}[!hbt]
	\centering
	\includegraphics[height=0.6\textwidth]{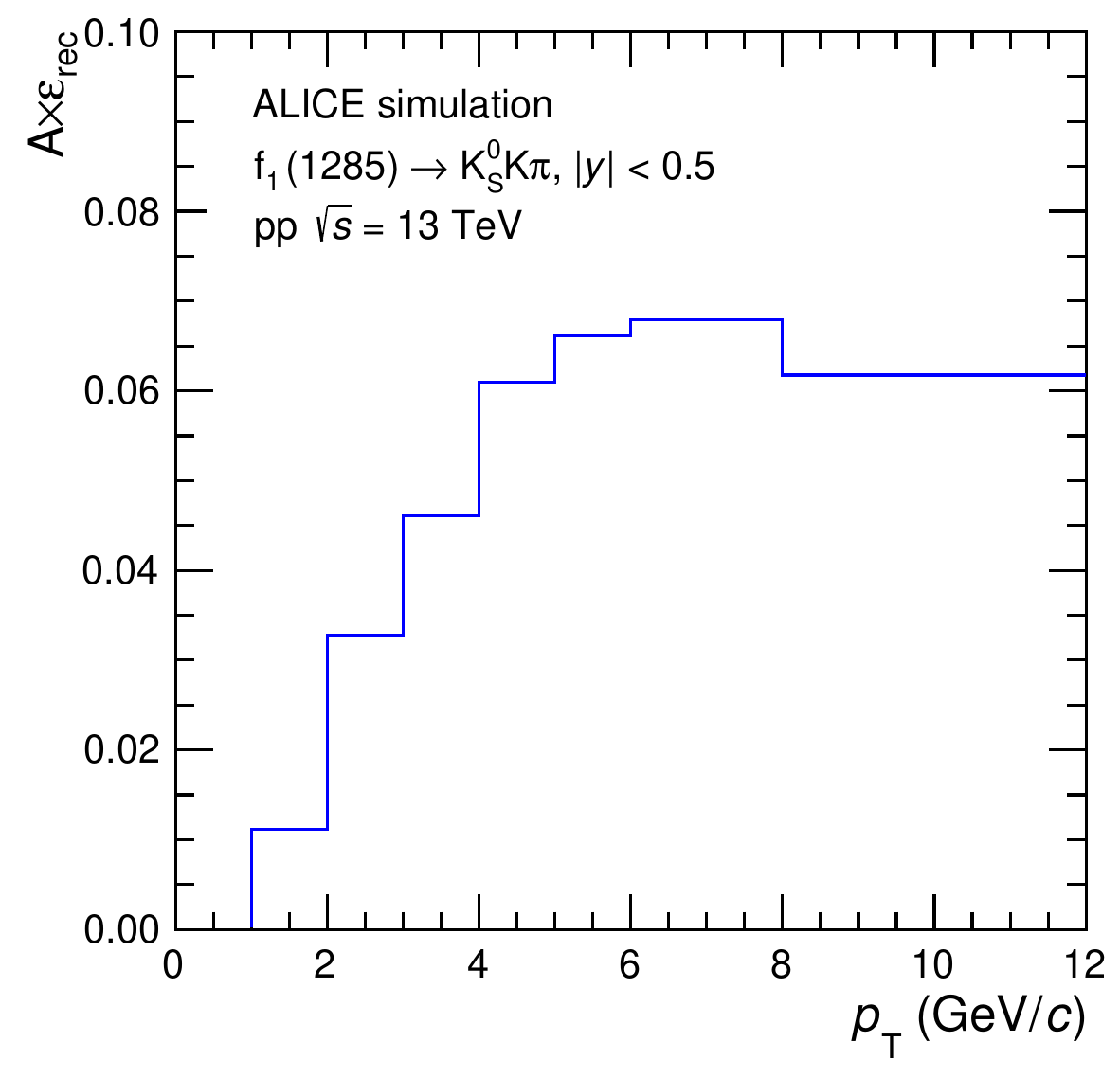}
	\caption{\label{fig:eff} The product of the acceptance and the resonance reconstruction efficiency as a function of \pt for \fone\xspace at midrapidity ($|y|<$ 0.5) in simulated pp collisions at \ENthirteen.}
\end{figure}

The extracted raw yields ($N^{\mathrm{raw}}$) are corrected for detector acceptance and reconstruction efficiency ($A\times\epsilon_{\mathrm{rec}}$) as well as the BR of the analyzed decay channel. The product $A\times\epsilon_{\mathrm{rec}}$ is estimated using simulated pp events produced with the PYTHIA8 Monte Carlo (MC) event generator~\cite{Skands:2014pea}, in which \fone\xspace particles are injected with a flat \pt distribution. The particles are then propagated through the ALICE detector using the GEANT3 transport code~\cite{Brun:1994aa}. The $A\times\epsilon_{\mathrm{rec}}$, defined as the ratio of reconstructed to generated \fone, is calculated as a function of \pt within $|y|<$ 0.5. The event and track selections used in the data analysis are also applied in the simulation. Notably, $A\times\epsilon_{\mathrm{rec}}$ initially increases with \pt, starting at around 1\% at \pt = 1.5 GeV/$c$ and reaching a maximum value of approximately 6.5\% at \pt $\approx$ 6~GeV/$c$ before decreasing again, as depicted in Fig.~\ref{fig:eff}. The relative statistical uncertainties on $A\times\epsilon_{\mathrm{rec}}$ are found to be in the range of 5--10\% across the \pt intervals. Moreover, since the generated \pt spectra of \fone\xspace have a different shape than the measured \pt spectra, a reweighting procedure~\cite{ALICE:2021ptz, ALICE:2020jsh} is implemented iteratively until convergence is reached by correcting at first the measured raw yields with the reconstruction efficiency obtained with the generated \pt spectra. The resulting \pt spectrum is then fitted with a Levy--Tsallis function and the extracted parametrization is finally used to weight the Monte Carlo spectra at generated and reconstructed level. From these reweighted spectra, the $A\times\epsilon_{\mathrm{rec}}$ as a function of \pt is determined.

The measurements need to be further corrected for trigger inefficiency ($\epsilon_{\rm trig}$), vertex reconstruction inefficiency ($\epsilon_{\rm vert}$), and signal loss ($f_{\rm SL}$) factors, which are determined through MC simulations. Signal loss factor accounts for the loss of \fone\xspace mesons due to trigger selection (i.e. \fone\xspace mesons produced in pp collisions that did not fire the trigger). Given the potential limitations of simulations involving injected \fone\xspace signals in realistically assessing correction factors~\cite{ALICE:2022qnb}, these factors are taken to be the same as for the K$^{*0}$ meson at the same collision energy~\cite{ALICE:2020jsh}. The values of the correction factors for pp collisions at \s = 13 TeV are $\epsilon_{\rm trig}$ = 0.74, $\epsilon_{\rm vert}$ = 0.93. The signal loss correction factor ($f_{\rm SL}$) is smaller than 2\% for \pt $>$ 1 GeV/$c$~\cite{ALICE:2020jsh}. Finally, the yields are normalized by the number of accepted events ($N_{\mathrm{event}}^{\rm{acc}}$) to obtain the \fone\xspace \pt-differential yield in inelastic pp collision, which can be formally expressed as

\begin{equation}
	\frac{1}{N_{\mathrm{evt}}} \frac{\mathrm{d}^{2}N}{\mathrm{d}y\mathrm{d}p_{\mathrm{T}}}= \frac{1}{N^{\rm{acc}}_{\rm{evt}}}\frac{N^{\mathrm{raw}}}{\Delta y \Delta p_{\mathrm{T}}}\frac{\epsilon_{\mathrm{trig}}\ \epsilon_{\mathrm{vert}}\ f_{\mathrm{SL}}}{(\mathrm{A}\times\epsilon_{\mathrm{rec}})\mathrm{BR}},
\end{equation}
where $\frac{\mathrm{d}^{2}N}{\mathrm{d}y\mathrm{d}p_{\mathrm{T}}}$ is the number of \fone\xspace produced in a given rapidity (d$y$) and transverse momentum (d\pt) interval. 

\section{Systematic uncertainties}\label{SU}

For the measurement of the \fone\xspace mass and yields, various sources of systematic uncertainties have been taken into account: the signal extraction method, the primary track selections, the \KS\xspace reconstruction and selection, the particle identification criteria, the method adopted in matching track segments in the ITS with tracks in the TPC, as well as uncertainties in the material budget and hadronic interactions of the produced particles in the ALICE detectors. The resulting changes in the \fone\xspace mass and yields for each \pt interval, obtained from repeating the entire analysis chain by varying one source at a time (as described below) while keeping others at default, are incorporated as systematic uncertainties. Table~\ref{tab:sysunc} summarizes the systematic uncertainties on the measured \fone\xspace yield and mass across the analyzed \pt range.

\begin{table}[ht]
	\centering
	\caption{Systematic uncertainties on measured \fone\xspace yield and mass in pp collisions at \ENthirteen.}
	\begin{tabular}{|c|c|c|}	
		\hline
		Systematic variation ($\%$) & \multicolumn{2}{|c|}{Quantity of interest}\\
		\hline
		& Yield & Mass\\
		\hline
		Signal extraction &  10.5--14.5  & 0.13--0.19 \\
		\hline
		Primary track selection & 3.9--5.9 & 0.03--0.08\\
		\hline
		Secondary track selection & 6.4--9.2 & 0.04--0.08\\
		\hline
		Particle identification & 1.0--6.5  & 0.003--0.027\\
		\hline
		ITS-TPC matching & 5.0  & -\\
		\hline
		Material budget & 1.8  & -\\
		\hline
		Hadronic interaction & 2.0  & -\\
		\hline
		Total & 16--17 & 0.16--0.20\\
		\hline	 
	\end{tabular}
	\label{tab:sysunc}	
\end{table}

Several factors are varied to evaluate the uncertainty in the signal extraction from the invariant mass fits, including fitting ranges, residual background fit function, and variations in the width of the three resonances (\fone, f$_{1}$(1420), and $\eta$(1475)). When adjusting fitting range boundaries, a shift of $\pm$ 20 MeV/$c^{2}$ with respect to the default case is applied to both sides. The widths of all resonances are treated as free parameters in the fit, unlike the default case where they are fixed to their world-average values, and the differences in \fone\xspace mass and yields contribute to the systematic uncertainties. Additionally, the residual background is modeled using second and third-order polynomials to investigate systematic deviations on the mass and yield of \fone. Moreover, the mass of f$_{1}$(1420) is held constant, unlike in the standard case where it is allowed to vary, to understand its impact on the observed \fone\xspace mass and yield. The resulting uncertainty for signal extraction on the observed \fone\xspace mass and yield varies from 0.13\% to 0.19\% and 10.5\% to 14.5\%, respectively, across the measured \pt ranges. 
For the primary-track selection, the criteria are varied following the procedure outlined in Ref.~\cite{ALICE:2020jsh}. This results in an uncertainty on the \fone\xspace mass ranging from 0.03\% to 0.08\% and an uncertainty on its yield ranging from 3.9\% to 5.9\% across the various \pt intervals. The uncertainty due to the \KS\xspace reconstruction is estimated by varying the selections in Table~\ref{tab:K0ssel}, resulting in a \pt-dependent systematic uncertainty ranging from 0.04\% to 0.08\% for the \fone\xspace mass and from 6.4\% to 9.2\% for the \fone\xspace yield. The uncertainties associated with the identification of the pions and kaons produced in the \fone\xspace decay are assessed by varying the selection criteria in the TOF from $|\mathrm{n} \sigma_{\mathrm{TOF}}| <$ 3 to $|\mathrm{n} \sigma_{\mathrm{TOF}}| <$ 4. This variation results in \fone\xspace mass uncertainties ranging from 0.003\% to 0.027\% and yield uncertainties ranging from 1\% to 6.5\%, depending on \pt. Furthermore, uncertainties related to the material budget, the cross section for hadronic interactions in the detector material, and the ITS--TPC matching efficiency, obtained from Ref.~\cite{ALICE:2020jsh}, contribute to the uncertainty on the yield of \fone. The total uncertainty is obtained by summing the uncertainties from all sources in quadrature. The uncertainty on the \fone\xspace mass ranges from approximately 0.16\% to 0.20\%, while for the yield, it spans from 16\% to 17\% across the measured \pt intervals.

\section{Results} \label{RE}
\begin{figure}[!hbt]
	\centering
	\includegraphics[height=0.6\textwidth]{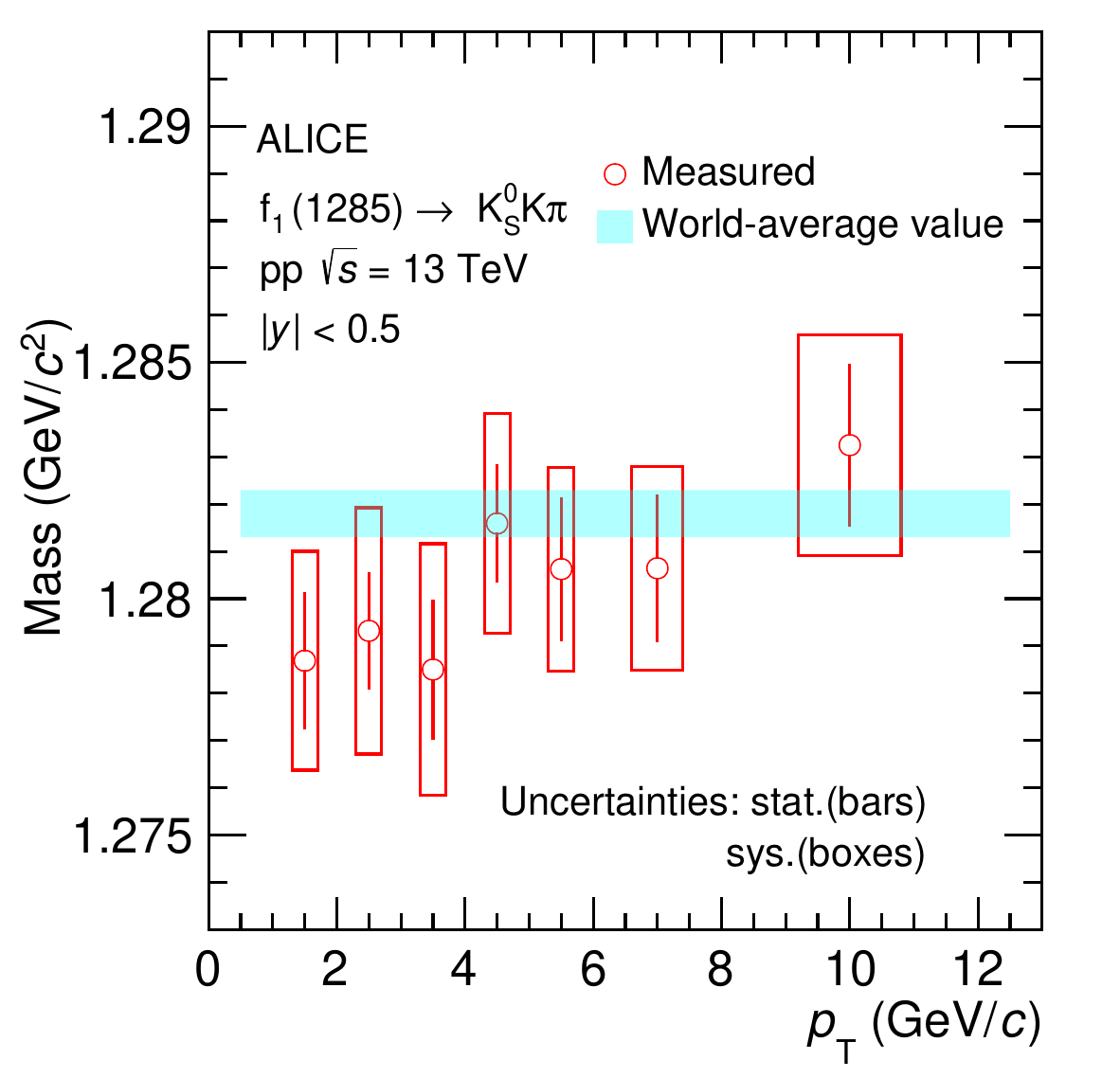}
	\caption{\label{fig:mass} Measured \fone\xspace mass as a function of \pt at midrapidity ($|y|<$ 0.5) in minimum-bias pp collisions at \ENthirteen. The statistical and systematic uncertainties are shown as bars and boxes, respectively. The blue band represents the world-average value for the mass of \fone\xspace~\cite{ParticleDataGroup:2022pth} having an uncertainty of 0.5 MeV/$c^{2}$.}
\end{figure}

The mass of \fone\xspace resonance, i.e., the fit parameter $M_{0}$ obtained from Eq.~\ref{eq:Fitfunction}, is shown in Fig.~\ref{fig:mass} for the different \pt intervals considered in this analysis. The systematic uncertainties on the measured mass, shown as boxes around the data points, are evaluated following the description in Sec.~\ref{SU}. The measured sample-average mass, 1.28 $\pm$ 0.001 GeV/c$^{2}$ is in excellent agreement with the world average value of 1.281 $\pm$ 0.0005 GeV/c$^{2}$ within uncertainties.

\begin{figure}[!hbt]
	\centering
	\includegraphics[height=0.6\textwidth]{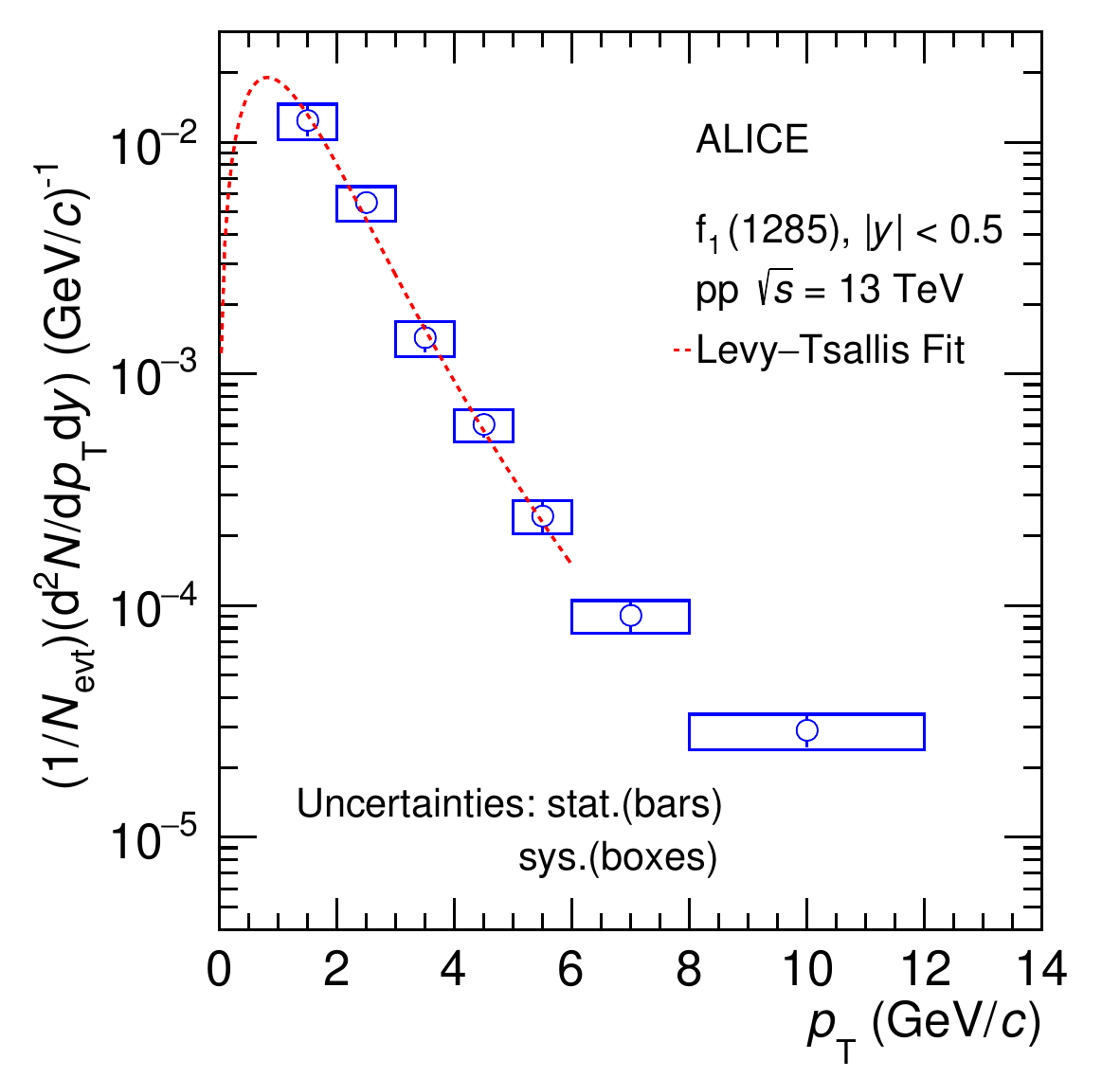}
	\caption{\label{fig:corrspectrapp} \pt-differential yield of \fone\xspace measured at midrapidity ($|y|<$ 0.5) in inelastic pp collisions at \ENthirteen. The statistical and systematic uncertainties are shown as bars and boxes, respectively. The data points are fitted using a Levy--Tsallis function~\cite{Tsallis:1987eu} and shown by the red dashed line. The BR uncertainty for \foneDC\xspace is 0.1\%.}
\end{figure}

Figure~\ref{fig:corrspectrapp} illustrates the \fone\xspace \pt-differential yield in pp collisions at \ENthirteen, incorporating all the corrections detailed in Sec.~\ref{SU}. The \pt spectrum is fitted with a Levy--Tsallis function, a combination of an exponential and power law function~\cite{Tsallis:1987eu}, to extrapolate the yield down to zero \pt. An exponential function describes the low-\pt section of the spectrum, while a power law characterizes the high-\pt region. Since there are only two \pt bins above 6 GeV/$c$ with large bin width, the Levy--Tsallis fit in the default case is performed in the 0 $<p_{\rm T}<$ 6 GeV/$c$ range. 

This fitting procedure enables the extraction of the \pt-integrated yield (d$N$/d$y$) and the average transverse momentum (\meanpt) of \fone, utilizing both the measured and extrapolated distributions. The extrapolation to the low-\pt ($<$ 1~GeV/$c$) region encompasses approximately 41\% of the total \fone\xspace yield. The high-\pt extrapolation is found to be negligible. The \meanpt is determined by evaluating the mean value of the fit function within each \pt bin, weighted by the measured yield in that bin. The systematic uncertainties in the \pt spectrum, arising from the various sources described in Sec.~\ref{SU}, contribute to the systematic uncertainties in d$N$/d$y$ and \meanpt. The systematic uncertainties due to the extrapolation are evaluated by varying the fit functions: the Boltzmann--Gibbs blast wave function~\cite{Schnedermann:1993ws}, Bose--Einstein distribution, and $m_{\rm T}$ exponential~\cite{ALICE:2020jsh} are considered in place of the Levy--Tsallis. The uncertainties of d$N$/d$y$ and \meanpt are approximately 31\% and 17\%, respectively. The dominant contribution to the \meanpt uncertainty arises from the low-\pt extrapolation ($\sim$14\%), estimated conservatively using the largest deviation from the Levy-Tsallis function. If the RMS of the \meanpt variations is used instead, the extrapolation uncertainty is reduced to $\sim$9\%.

\begin{figure}[!hbt]
 	\centering
 	\includegraphics[height=0.6\textwidth]{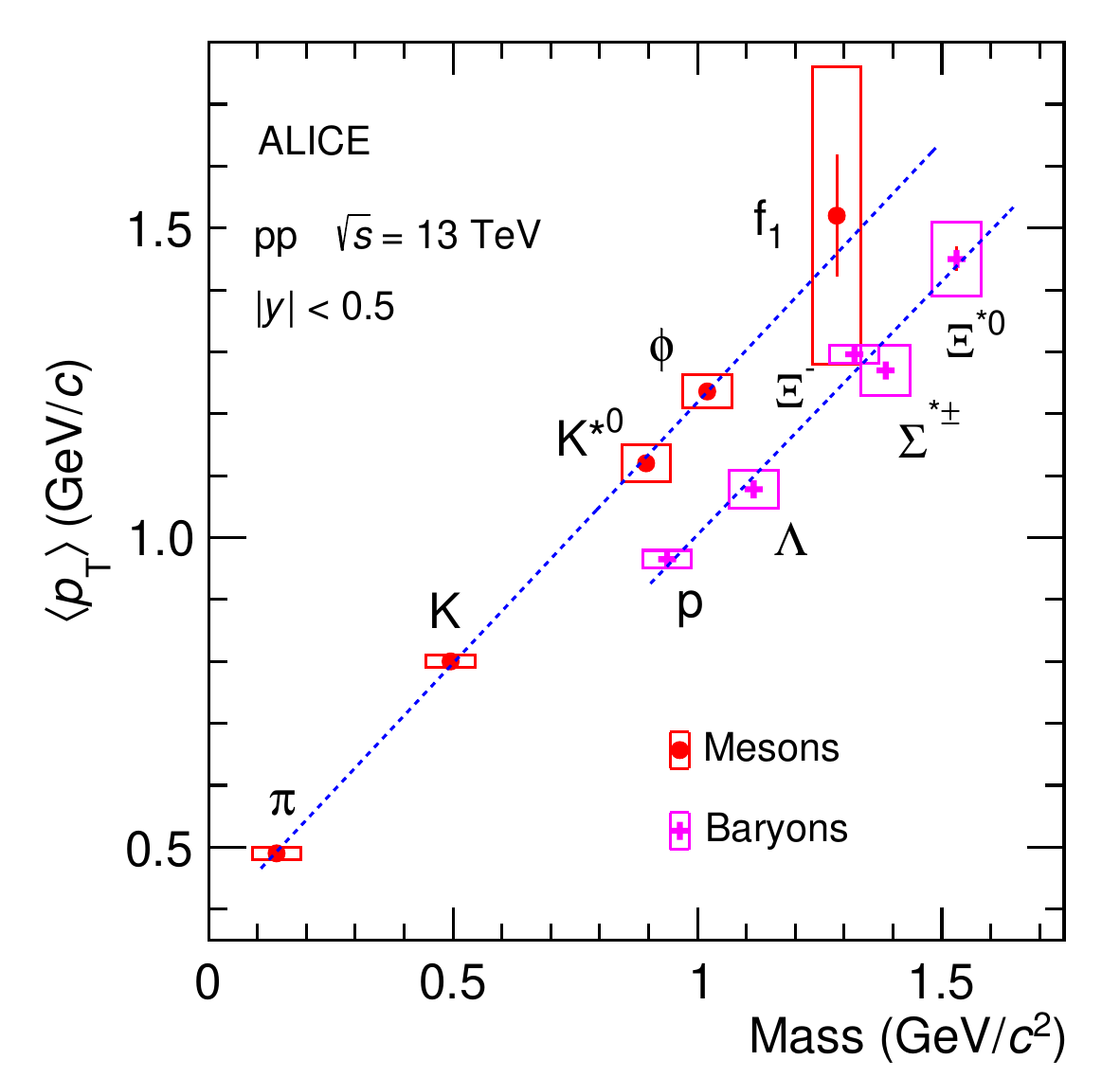}
 	\caption{\label{fig:mptvsmass} Average transverse momentum of light-flavor hadrons as a function of hadron mass at midrapidity ($|y|<$ 0.5) in inelastic pp collisions at \ENthirteen. The statistical and systematic uncertainties are shown as bars and boxes, respectively. The blue dotted lines are linear fits to the data points.}
\end{figure}

Table~\ref{tab:TM} shows d$N$/d$y$ and \meanpt and their uncertainties in inelastic pp collisions at \s = 13 TeV. Figure~\ref{fig:mptvsmass} compares the average transverse momentum of \fone\xspace with that of all other light-flavor hadrons~\cite{ALICE:2020jsh, ALICE:2023egx} measured at midrapidity ($|y|<$ 0.5) in pp collisions at \ENthirteen. Two distinct linear trends are observed, one for mesons and the other for baryons. For particles with similar masses (K$^{*0}$, p, $\phi$, $\Lambda$, \foneshort, $\Xi^{-}$), mesons exhibit a higher average transverse momentum than baryons. Notably, \fone\xspace aligns with the linear trend of other mesons, although with large conservative systematic uncertainties. This observation suggests that \fone\xspace may have an ordinary meson structure.

The \pt-integrated yield is further compared with calculations from the canonical-ensemble-based statistical hadronization model ($\gamma_{\rm S}$CSM)~\cite{Vovchenko:2019kes}, also shown in Table~\ref{tab:TM}. The conventional statistical framework employs an ideal hadron--resonance gas (HRG) in thermal and chemical equilibrium at the chemical freeze-out stage. In the canonical ensemble, the values of three Abelian charges --- baryon number ($B$), electric charge ($Q$), and strangeness ($S$) --- are fixed and conserved exactly across the designated correlation volume $V_{\rm C}$. In this model, the multiplicity dependence of hadron production is influenced by the canonical suppression of these three Abelian charges. It incorporates the incomplete equilibrium of strangeness via the strangeness saturation parameter $\gamma_{\rm S}$ and effectively reproduces various multiplicity-dependent hadron-to-pion ratios~\cite{Vovchenko:2019kes}. Thermal fits to the yields of various particles, including $\pi$, K, p, K$^{*0}$, $\Lambda$, $\Omega$, K$^{0}_{\rm S}$, $\Xi$, and $\phi$, as measured by the ALICE Collaboration in pp collisions at \ENthirteen~\cite{ALICE:2020jsh}, have been conducted. The fit parameters include the freeze-out temperature, radius of the produced fireball, $V_{\rm C}$, and $\gamma_{\rm S}$. It is assumed that the baryon chemical potential is zero~\cite{ALICE:2023ulv}. The thermal model calculations for the \pt-integrated yield of \fone\xspace are carried out for two different scenarios: The first scenario assumes |$S$| = 0, indicating that the \fone\xspace meson does not contain any valence strange or anti-strange quarks. The second scenario considers |$S$| = 2, which corresponds to the presence of one strange and one anti-strange quark within the \fone\xspace meson. The calculated yield with |$S$| = 0 scenario is consistent with the experimental measurement.

\begin{table}[!hbt]
	\centering
	\caption{The \pt-integrated yield and average transverse momentum of the \fone\xspace meson in proton--proton collisions at center-of-mass energy of 13 TeV. The comparison of the \pt-integrated yield of \fone\xspace from ALICE data with thermal model ($\gamma_{\rm S}$CSM) calculations~\cite{Vovchenko:2019kes} is shown.}
	\begin{tabular}{llll}	
		\hline
		& ALICE data & \multicolumn{2}{l}{Thermal model}\\
		\hline
		&  & |$S$|=0 & |$S$|=2\\
		d$N$/d$y$ &  0.034 $\pm$ 0.004 (stat) $\pm$ 0.010 (sys)  & 0.025 &  0.014 \\
		\meanpt (GeV/$c$) & 1.52 $\pm$ 0.10 (stat) $\pm$ 0.24 (sys) & - & -\\
     \hline
	\end{tabular}
	\label{tab:TM}	
\end{table}

\begin{figure}[!hbt]
	\centering
	\includegraphics[height=0.6\textwidth]{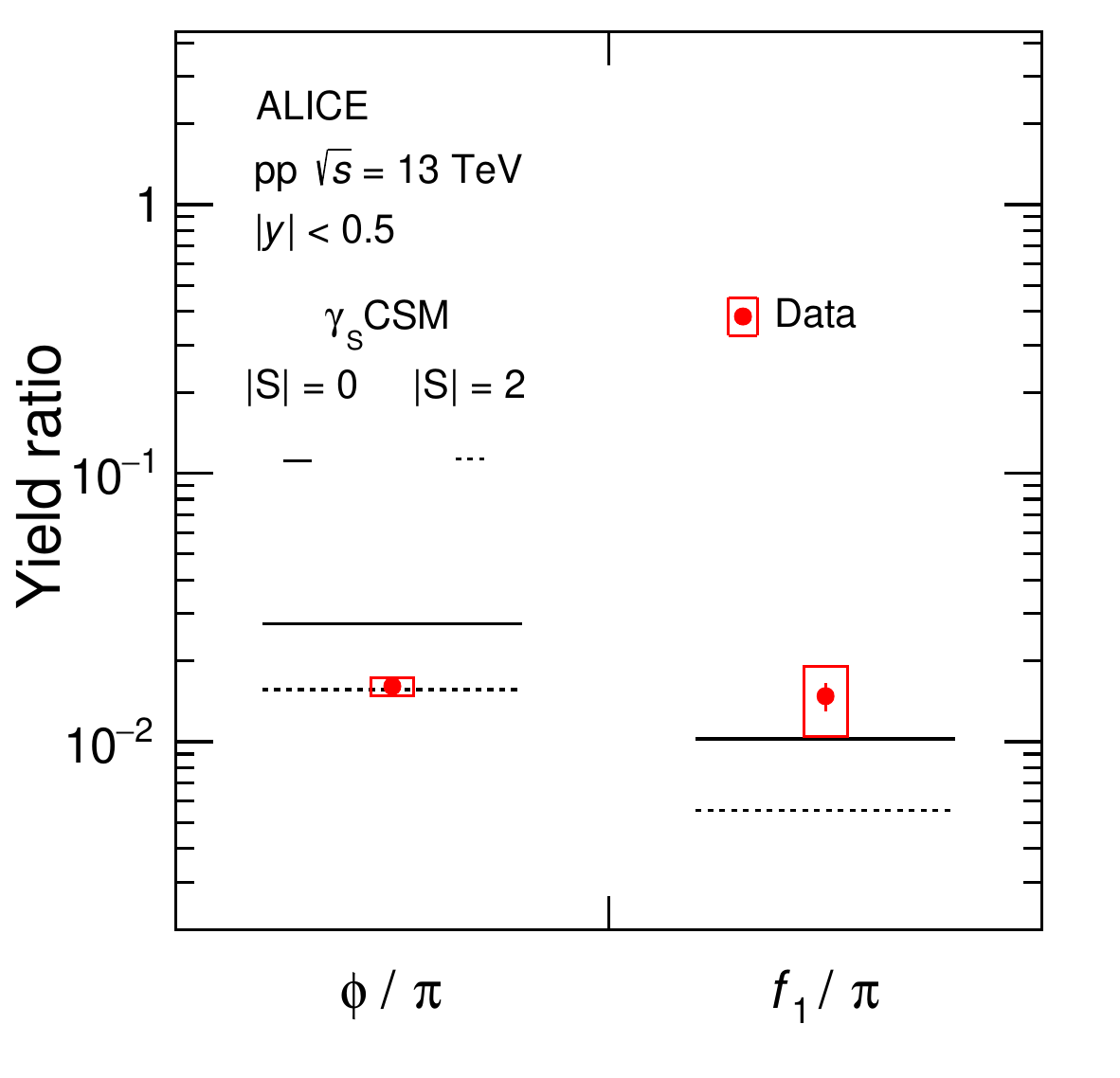}
	\caption{\label{fig:qc} The transverse-momentum-integrated yield ratio of $\phi$/$\pi$ (left)~\cite{ALICE:2020jsh} and \foneshort/$\pi$ (right) measured in inelastic pp collisions at \ENthirteen. The statistical and systematic uncertainties on the data points are shown as bars and boxes, respectively. The black solid and dotted lines represent the calculations from the $\gamma_{\rm S}$CSM with different strangeness content of $\phi$ and \foneshort\xspace mesons.}
\end{figure}

To gain insights into the valence quark composition of the \fone\xspace meson, the \pt-integrated yield ratio of \foneshort/$\pi$ in pp collisions at \ENthirteen is compared with calculations from the $\gamma_{\rm S}$CSM, as depicted in Fig.~\ref{fig:qc}. At first, as a baseline check for this methodology, the $\phi$/$\pi$ ratio is calculated by $\gamma_{\rm S}$CSM with two scenarios and
compared with experimental data~\cite{ALICE:2020jsh}. The $\phi$ meson is a neutral particle comprising a strange quark--antiquark pair. It has a net strangeness of zero, thus remaining unaffected by the precise conservation of strangeness in the canonical suppression picture. However, in the strangeness nonequilibrium picture, the $\phi$ meson is considered a double-strange particle (|$S$| = 2)~\cite{Vovchenko:2019kes}. Thus, the experimental data is compared in Fig.~\ref{fig:qc} with $\gamma_{\rm S}$CSM calculations for |$S$| = 0 (indicating a total strangeness content of $\phi$ to be zero, depicted by the solid line) and |$S$| = 2 (indicating a hidden strangeness content of $\phi$ to be two, represented by the dotted line). As expected in the strangeness nonequilibrium picture, the $\gamma_{\rm S}$CSM calculation for $\phi$/$\pi$ ratio with the $\phi$ meson having |$S$| = 0 shows a large deviation of 9.15$\sigma$ from the experimental measurements, whereas |$S$| = 2 is in good agreement with the experimental measurements within 0.5$\sigma$. \\
The calculation of the \foneshort/$\pi$ ratio from $\gamma_{\rm S}$CSM is carried out for the two different scenarios of |$S$| = 0 (represented by the solid line) and |$S$| = 2 (represented by the dotted line). The measured \foneshort/$\pi$ ratio deviates by 0.96$\sigma$ from |$S$| = 0 and by 1.97$\sigma$ from |$S$| = 2, indicating that the $\gamma_{\rm S}$CSM calculation with |$S$| = 0 is favored over |$S$| = 2 by the ALICE data. This observation is further validated through a $\chi^{2}$ statistical hypothesis test, which quantifies the difference between experimental data and model predictions. The goodness-of-fit is assessed using the $\chi^{2}$ value, which is then converted into a right-tailed $p$-value. From standard $\chi^{2}$ probability tables, the right-tailed $p$-value can be obtained for a given number of degrees of freedom. A common criterion for statistical significance is $p < 0.05$, indicating a meaningful deviation between the data and the model expectations. In this analysis, the $p$-value for $|S| = 0$ is found to be 0.33, suggesting no significant discrepancy between the data and the model. However, for $|S| = 2$, the $p$-value is 0.04, which corresponds to a confidence level of $1 - 0.04 = 96\%$. This implies that with 96\% confidence, the data significantly deviates from the model predictions under the $|S| = 2$ hypothesis.\\
Therefore, this study suggests that \fone\xspace is more likely to have no strange quark content than a combination of a strange and an anti-strange quark. This finding contradicts the hypothesis that \fone\xspace is a tetraquark state (according to the $\gamma_{\rm S}$CSM model) and is consistent with the results of the LHCb Collaboration~\cite{LHCb:2013ged}.

\section{Summary} \label{Summ}

The ALICE Collaboration presents the first measurement of the \fone\xspace meson production in inelastic proton--proton collisions at \s = 13 TeV. This measurement spans a wide transverse momentum range from 1 to 12 GeV/$c$ at midrapidity ($|y|<$ 0.5). The mass of \fone\xspace reconstructed from the \KS K$^{\pm} \pi^{\mp}$ decays is in good agreement with the world-average value within the uncertainties. Notably, the average transverse momentum of \fone\xspace aligns with the linear trend with mass observed for other mesons and it is higher, although compatible within 1$\sigma$ of the systematic uncertainty, with the \meanpt of baryons of similar masses. Moreover, the $\gamma_{\rm S}$CSM of the \foneshort/$\pi$ \pt-integrated yield ratio, considering no strange quarks inside \fone, agrees with the ALICE data within 1$\sigma$. However, it deviates by $\sim$2$\sigma$ when assuming the presence of one strange and one anti-strange quark. These observations suggest that the state of \fone\xspace is a conventional meson, which disfavors the tetraquark hypothesis and aligns with the findings of the LHCb Collaboration. With larger data samples available in Run 3 and Run 4, combined with the improved tracking efficiency of the upgraded ITS detector, it may become feasible to reconstruct the \fone\xspace meson at low transverse momentum ($<$ 1 GeV/$c$), thereby improving the significance of future analyses. Additionally, future studies of the elliptic flow of \fone\xspace and femtoscopy measurements in the K$^{*}\overline{\rm K}$ coupled channel, using the large data samples from Run 3 and the upcoming Run 4, may help to distinguish the di-quark or molecular nature of \fone.



\newenvironment{acknowledgement}{\relax}{\relax}
\begin{acknowledgement}
\section*{Acknowledgements}

The ALICE Collaboration would like to thank all its engineers and technicians for their invaluable contributions to the construction of the experiment and the CERN accelerator teams for the outstanding performance of the LHC complex.
The ALICE Collaboration gratefully acknowledges the resources and support provided by all Grid centres and the Worldwide LHC Computing Grid (WLCG) collaboration.
The ALICE Collaboration acknowledges the following funding agencies for their support in building and running the ALICE detector:
A. I. Alikhanyan National Science Laboratory (Yerevan Physics Institute) Foundation (ANSL), State Committee of Science and World Federation of Scientists (WFS), Armenia;
Austrian Academy of Sciences, Austrian Science Fund (FWF): [M 2467-N36] and Nationalstiftung f\"{u}r Forschung, Technologie und Entwicklung, Austria;
Ministry of Communications and High Technologies, National Nuclear Research Center, Azerbaijan;
Conselho Nacional de Desenvolvimento Cient\'{\i}fico e Tecnol\'{o}gico (CNPq), Financiadora de Estudos e Projetos (Finep), Funda\c{c}\~{a}o de Amparo \`{a} Pesquisa do Estado de S\~{a}o Paulo (FAPESP) and Universidade Federal do Rio Grande do Sul (UFRGS), Brazil;
Bulgarian Ministry of Education and Science, within the National Roadmap for Research Infrastructures 2020-2027 (object CERN), Bulgaria;
Ministry of Education of China (MOEC) , Ministry of Science \& Technology of China (MSTC) and National Natural Science Foundation of China (NSFC), China;
Ministry of Science and Education and Croatian Science Foundation, Croatia;
Centro de Aplicaciones Tecnol\'{o}gicas y Desarrollo Nuclear (CEADEN), Cubaenerg\'{\i}a, Cuba;
Ministry of Education, Youth and Sports of the Czech Republic, Czech Republic;
The Danish Council for Independent Research | Natural Sciences, the VILLUM FONDEN and Danish National Research Foundation (DNRF), Denmark;
Helsinki Institute of Physics (HIP), Finland;
Commissariat \`{a} l'Energie Atomique (CEA) and Institut National de Physique Nucl\'{e}aire et de Physique des Particules (IN2P3) and Centre National de la Recherche Scientifique (CNRS), France;
Bundesministerium f\"{u}r Bildung und Forschung (BMBF) and GSI Helmholtzzentrum f\"{u}r Schwerionenforschung GmbH, Germany;
General Secretariat for Research and Technology, Ministry of Education, Research and Religions, Greece;
National Research, Development and Innovation Office, Hungary;
Department of Atomic Energy Government of India (DAE), Department of Science and Technology, Government of India (DST), University Grants Commission, Government of India (UGC) and Council of Scientific and Industrial Research (CSIR), India;
National Research and Innovation Agency - BRIN, Indonesia;
Istituto Nazionale di Fisica Nucleare (INFN), Italy;
Japanese Ministry of Education, Culture, Sports, Science and Technology (MEXT) and Japan Society for the Promotion of Science (JSPS) KAKENHI, Japan;
Consejo Nacional de Ciencia (CONACYT) y Tecnolog\'{i}a, through Fondo de Cooperaci\'{o}n Internacional en Ciencia y Tecnolog\'{i}a (FONCICYT) and Direcci\'{o}n General de Asuntos del Personal Academico (DGAPA), Mexico;
Nederlandse Organisatie voor Wetenschappelijk Onderzoek (NWO), Netherlands;
The Research Council of Norway, Norway;
Pontificia Universidad Cat\'{o}lica del Per\'{u}, Peru;
Ministry of Science and Higher Education, National Science Centre and WUT ID-UB, Poland;
Korea Institute of Science and Technology Information and National Research Foundation of Korea (NRF), Republic of Korea;
Ministry of Education and Scientific Research, Institute of Atomic Physics, Ministry of Research and Innovation and Institute of Atomic Physics and Universitatea Nationala de Stiinta si Tehnologie Politehnica Bucuresti, Romania;
Ministry of Education, Science, Research and Sport of the Slovak Republic, Slovakia;
National Research Foundation of South Africa, South Africa;
Swedish Research Council (VR) and Knut \& Alice Wallenberg Foundation (KAW), Sweden;
European Organization for Nuclear Research, Switzerland;
Suranaree University of Technology (SUT), National Science and Technology Development Agency (NSTDA) and National Science, Research and Innovation Fund (NSRF via PMU-B B05F650021), Thailand;
Turkish Energy, Nuclear and Mineral Research Agency (TENMAK), Turkey;
National Academy of  Sciences of Ukraine, Ukraine;
Science and Technology Facilities Council (STFC), United Kingdom;
National Science Foundation of the United States of America (NSF) and United States Department of Energy, Office of Nuclear Physics (DOE NP), United States of America.
In addition, individual groups or members have received support from:
Czech Science Foundation (grant no. 23-07499S), Czech Republic;
FORTE project, reg.\ no.\ CZ.02.01.01/00/22\_008/0004632, Czech Republic, co-funded by the European Union, Czech Republic;
European Research Council (grant no. 950692), European Union;
ICSC - Centro Nazionale di Ricerca in High Performance Computing, Big Data and Quantum Computing, European Union - NextGenerationEU;
Academy of Finland (Center of Excellence in Quark Matter) (grant nos. 346327, 346328), Finland.

\end{acknowledgement}

\bibliographystyle{utphys}   
\bibliography{bibliographylatest}

\newpage
\appendix

%
%

\section{The ALICE Collaboration}
\label{app:collab}
\begin{flushleft} 
\small

S.~Acharya\,\orcidlink{0000-0002-9213-5329}\,$^{\rm 127}$, 
A.~Agarwal$^{\rm 135}$, 
G.~Aglieri Rinella\,\orcidlink{0000-0002-9611-3696}\,$^{\rm 32}$, 
L.~Aglietta\,\orcidlink{0009-0003-0763-6802}\,$^{\rm 24}$, 
M.~Agnello\,\orcidlink{0000-0002-0760-5075}\,$^{\rm 29}$, 
N.~Agrawal\,\orcidlink{0000-0003-0348-9836}\,$^{\rm 25}$, 
Z.~Ahammed\,\orcidlink{0000-0001-5241-7412}\,$^{\rm 135}$, 
S.~Ahmad\,\orcidlink{0000-0003-0497-5705}\,$^{\rm 15}$, 
S.U.~Ahn\,\orcidlink{0000-0001-8847-489X}\,$^{\rm 71}$, 
I.~Ahuja\,\orcidlink{0000-0002-4417-1392}\,$^{\rm 37}$
% ...

\section*{Affiliation Notes}

$^{\rm I}$ Deceased\\
$^{\rm II}$ Also at: Max-Planck-Institut fur Physik, Munich, Germany\\
$^{\rm III}$ Also at: Italian National Agency for New Technologies, Energy and Sustainable Economic Development (ENEA), Bologna, Italy\\
$^{\rm IV}$ Also at: Dipartimento DET del Politecnico di Torino, Turin, Italy\\
$^{\rm V}$ Also at: Yildiz Technical University, Istanbul, T\"{u}rkiye\\
$^{\rm VI}$ Also at: Department of Applied Physics, Aligarh Muslim University, Aligarh, India\\
$^{\rm VII}$ Also at: Institute of Theoretical Physics, University of Wroclaw, Poland\\
$^{\rm VIII}$ Also at: An institution covered by a cooperation agreement with CERN\\

\section*{Collaboration Institutes}

$^{1}$ A.I. Alikhanyan National Science Laboratory (Yerevan Physics Institute) Foundation, Yerevan, Armenia\\
$^{2}$ AGH University of Krakow, Cracow, Poland\\
$^{3}$ Bogolyubov Institute for Theoretical Physics, National Academy of Sciences of Ukraine, Kiev, Ukraine\\
$^{4}$ Bose Institute, Department of Physics  and Centre for Astroparticle Physics and Space Science (CAPSS), Kolkata, India\\
$^{5}$ California Polytechnic State University, San Luis Obispo, California, United States\\
$^{6}$ Central China Normal University, Wuhan, China\\
$^{7}$ Centro de Aplicaciones Tecnol\'{o}gicas y Desarrollo Nuclear (CEADEN), Havana, Cuba\\
$^{8}$ Centro de Investigaci\'{o}n y de Estudios Avanzados (CINVESTAV), Mexico City and M\'{e}rida, Mexico\\
$^{9}$ Chicago State University, Chicago, Illinois, United States\\
$^{10}$ China Institute of Atomic Energy, Beijing, China\\
$^{11}$ China University of Geosciences, Wuhan, China\\
$^{12}$ Chungbuk National University, Cheongju, Republic of Korea\\
$^{13}$ Comenius University Bratislava, Faculty of Mathematics, Physics and Informatics, Bratislava, Slovak Republic\\
$^{14}$ Creighton University, Omaha, Nebraska, United States\\
$^{15}$ Department of Physics, Aligarh Muslim University, Aligarh, India\\
$^{16}$ Department of Physics, Pusan National University, Pusan, Republic of Korea\\
$^{17}$ Department of Physics, Sejong University, Seoul, Republic of Korea\\
$^{18}$ Department of Physics, University of California, Berkeley, California, United States\\
$^{19}$ Department of Physics, University of Oslo, Oslo, Norway\\
$^{20}$ Department of Physics and Technology, University of Bergen, Bergen, Norway\\
$^{21}$ Dipartimento di Fisica, Universit\`{a} di Pavia, Pavia, Italy\\
$^{22}$ Dipartimento di Fisica dell'Universit\`{a} and Sezione INFN, Cagliari, Italy\\
$^{23}$ Dipartimento di Fisica dell'Universit\`{a} and Sezione INFN, Trieste, Italy\\
$^{24}$ Dipartimento di Fisica dell'Universit\`{a} and Sezione INFN, Turin, Italy\\
$^{25}$ Dipartimento di Fisica e Astronomia dell'Universit\`{a} and Sezione INFN, Bologna, Italy\\
$^{26}$ Dipartimento di Fisica e Astronomia dell'Universit\`{a} and Sezione INFN, Catania, Italy\\
$^{27}$ Dipartimento di Fisica e Astronomia dell'Universit\`{a} and Sezione INFN, Padova, Italy\\
$^{28}$ Dipartimento di Fisica `E.R.~Caianiello' dell'Universit\`{a} and Gruppo Collegato INFN, Salerno, Italy\\
$^{29}$ Dipartimento DISAT del Politecnico and Sezione INFN, Turin, Italy\\
$^{30}$ Dipartimento di Scienze MIFT, Universit\`{a} di Messina, Messina, Italy\\
$^{31}$ Dipartimento Interateneo di Fisica `M.~Merlin' and Sezione INFN, Bari, Italy\\
$^{32}$ European Organization for Nuclear Research (CERN), Geneva, Switzerland\\
$^{33}$ Faculty of Electrical Engineering, Mechanical Engineering and Naval Architecture, University of Split, Split, Croatia\\
$^{34}$ Faculty of Engineering and Science, Western Norway University of Applied Sciences, Bergen, Norway\\
$^{35}$ Faculty of Nuclear Sciences and Physical Engineering, Czech Technical University in Prague, Prague, Czech Republic\\
$^{36}$ Faculty of Physics, Sofia University, Sofia, Bulgaria\\
$^{37}$ Faculty of Science, P.J.~\v{S}af\'{a}rik University, Ko\v{s}ice, Slovak Republic\\
$^{38}$ Frankfurt Institute for Advanced Studies, Johann Wolfgang Goethe-Universit\"{a}t Frankfurt, Frankfurt, Germany\\
$^{39}$ Fudan University, Shanghai, China\\
$^{40}$ Gangneung-Wonju National University, Gangneung, Republic of Korea\\
$^{41}$ Gauhati University, Department of Physics, Guwahati, India\\
$^{42}$ Helmholtz-Institut f\"{u}r Strahlen- und Kernphysik, Rheinische Friedrich-Wilhelms-Universit\"{a}t Bonn, Bonn, Germany\\
$^{43}$ Helsinki Institute of Physics (HIP), Helsinki, Finland\\
$^{44}$ High Energy Physics Group,  Universidad Aut\'{o}noma de Puebla, Puebla, Mexico\\
$^{45}$ Horia Hulubei National Institute of Physics and Nuclear Engineering, Bucharest, Romania\\
$^{46}$ HUN-REN Wigner Research Centre for Physics, Budapest, Hungary\\
$^{47}$ Indian Institute of Technology Bombay (IIT), Mumbai, India\\
$^{48}$ Indian Institute of Technology Indore, Indore, India\\
$^{49}$ INFN, Laboratori Nazionali di Frascati, Frascati, Italy\\
$^{50}$ INFN, Sezione di Bari, Bari, Italy\\
$^{51}$ INFN, Sezione di Bologna, Bologna, Italy\\
$^{52}$ INFN, Sezione di Cagliari, Cagliari, Italy\\
$^{53}$ INFN, Sezione di Catania, Catania, Italy\\
$^{54}$ INFN, Sezione di Padova, Padova, Italy\\
$^{55}$ INFN, Sezione di Pavia, Pavia, Italy\\
$^{56}$ INFN, Sezione di Torino, Turin, Italy\\
$^{57}$ INFN, Sezione di Trieste, Trieste, Italy\\
$^{58}$ Inha University, Incheon, Republic of Korea\\
$^{59}$ Institute for Gravitational and Subatomic Physics (GRASP), Utrecht University/Nikhef, Utrecht, Netherlands\\
$^{60}$ Institute of Experimental Physics, Slovak Academy of Sciences, Ko\v{s}ice, Slovak Republic\\
$^{61}$ Institute of Physics, Homi Bhabha National Institute, Bhubaneswar, India\\
$^{62}$ Institute of Physics of the Czech Academy of Sciences, Prague, Czech Republic\\
$^{63}$ Institute of Space Science (ISS), Bucharest, Romania\\
$^{64}$ Institut f\"{u}r Kernphysik, Johann Wolfgang Goethe-Universit\"{a}t Frankfurt, Frankfurt, Germany\\
$^{65}$ Instituto de Ciencias Nucleares, Universidad Nacional Aut\'{o}noma de M\'{e}xico, Mexico City, Mexico\\
$^{66}$ Instituto de F\'{i}sica, Universidade Federal do Rio Grande do Sul (UFRGS), Porto Alegre, Brazil\\
$^{67}$ Instituto de F\'{\i}sica, Universidad Nacional Aut\'{o}noma de M\'{e}xico, Mexico City, Mexico\\
$^{68}$ iThemba LABS, National Research Foundation, Somerset West, South Africa\\
$^{69}$ Jeonbuk National University, Jeonju, Republic of Korea\\
$^{70}$ Johann-Wolfgang-Goethe Universit\"{a}t Frankfurt Institut f\"{u}r Informatik, Fachbereich Informatik und Mathematik, Frankfurt, Germany\\
$^{71}$ Korea Institute of Science and Technology Information, Daejeon, Republic of Korea\\
$^{72}$ KTO Karatay University, Konya, Turkey\\
$^{73}$ Laboratoire de Physique Subatomique et de Cosmologie, Universit\'{e} Grenoble-Alpes, CNRS-IN2P3, Grenoble, France\\
$^{74}$ Lawrence Berkeley National Laboratory, Berkeley, California, United States\\
$^{75}$ Lund University Department of Physics, Division of Particle Physics, Lund, Sweden\\
$^{76}$ Nagasaki Institute of Applied Science, Nagasaki, Japan\\
$^{77}$ Nara Women{'}s University (NWU), Nara, Japan\\
$^{78}$ National and Kapodistrian University of Athens, School of Science, Department of Physics , Athens, Greece\\
$^{79}$ National Centre for Nuclear Research, Warsaw, Poland\\
$^{80}$ National Institute of Science Education and Research, Homi Bhabha National Institute, Jatni, India\\
$^{81}$ National Nuclear Research Center, Baku, Azerbaijan\\
$^{82}$ National Research and Innovation Agency - BRIN, Jakarta, Indonesia\\
$^{83}$ Niels Bohr Institute, University of Copenhagen, Copenhagen, Denmark\\
$^{84}$ Nikhef, National institute for subatomic physics, Amsterdam, Netherlands\\
$^{85}$ Nuclear Physics Group, STFC Daresbury Laboratory, Daresbury, United Kingdom\\
$^{86}$ Nuclear Physics Institute of the Czech Academy of Sciences, Husinec-\v{R}e\v{z}, Czech Republic\\
$^{87}$ Oak Ridge National Laboratory, Oak Ridge, Tennessee, United States\\
$^{88}$ Ohio State University, Columbus, Ohio, United States\\
$^{89}$ Physics department, Faculty of science, University of Zagreb, Zagreb, Croatia\\
$^{90}$ Physics Department, Panjab University, Chandigarh, India\\
$^{91}$ Physics Department, University of Jammu, Jammu, India\\
$^{92}$ Physics Program and International Institute for Sustainability with Knotted Chiral Meta Matter (SKCM2), Hiroshima University, Hiroshima, Japan\\
$^{93}$ Physikalisches Institut, Eberhard-Karls-Universit\"{a}t T\"{u}bingen, T\"{u}bingen, Germany\\
$^{94}$ Physikalisches Institut, Ruprecht-Karls-Universit\"{a}t Heidelberg, Heidelberg, Germany\\
$^{95}$ Physik Department, Technische Universit\"{a}t M\"{u}nchen, Munich, Germany\\
$^{96}$ Politecnico di Bari and Sezione INFN, Bari, Italy\\
$^{97}$ Research Division and ExtreMe Matter Institute EMMI, GSI Helmholtzzentrum f\"ur Schwerionenforschung GmbH, Darmstadt, Germany\\
$^{98}$ Saga University, Saga, Japan\\
$^{99}$ Saha Institute of Nuclear Physics, Homi Bhabha National Institute, Kolkata, India\\
$^{100}$ School of Physics and Astronomy, University of Birmingham, Birmingham, United Kingdom\\
$^{101}$ Secci\'{o}n F\'{\i}sica, Departamento de Ciencias, Pontificia Universidad Cat\'{o}lica del Per\'{u}, Lima, Peru\\
$^{102}$ Stefan Meyer Institut f\"{u}r Subatomare Physik (SMI), Vienna, Austria\\
$^{103}$ SUBATECH, IMT Atlantique, Nantes Universit\'{e}, CNRS-IN2P3, Nantes, France\\
$^{104}$ Sungkyunkwan University, Suwon City, Republic of Korea\\
$^{105}$ Suranaree University of Technology, Nakhon Ratchasima, Thailand\\
$^{106}$ Technical University of Ko\v{s}ice, Ko\v{s}ice, Slovak Republic\\
$^{107}$ The Henryk Niewodniczanski Institute of Nuclear Physics, Polish Academy of Sciences, Cracow, Poland\\
$^{108}$ The University of Texas at Austin, Austin, Texas, United States\\
$^{109}$ Universidad Aut\'{o}noma de Sinaloa, Culiac\'{a}n, Mexico\\
$^{110}$ Universidade de S\~{a}o Paulo (USP), S\~{a}o Paulo, Brazil\\
$^{111}$ Universidade Estadual de Campinas (UNICAMP), Campinas, Brazil\\
$^{112}$ Universidade Federal do ABC, Santo Andre, Brazil\\
$^{113}$ Universitatea Nationala de Stiinta si Tehnologie Politehnica Bucuresti, Bucharest, Romania\\
$^{114}$ University of Cape Town, Cape Town, South Africa\\
$^{115}$ University of Derby, Derby, United Kingdom\\
$^{116}$ University of Houston, Houston, Texas, United States\\
$^{117}$ University of Jyv\"{a}skyl\"{a}, Jyv\"{a}skyl\"{a}, Finland\\
$^{118}$ University of Kansas, Lawrence, Kansas, United States\\
$^{119}$ University of Liverpool, Liverpool, United Kingdom\\
$^{120}$ University of Science and Technology of China, Hefei, China\\
$^{121}$ University of South-Eastern Norway, Kongsberg, Norway\\
$^{122}$ University of Tennessee, Knoxville, Tennessee, United States\\
$^{123}$ University of the Witwatersrand, Johannesburg, South Africa\\
$^{124}$ University of Tokyo, Tokyo, Japan\\
$^{125}$ University of Tsukuba, Tsukuba, Japan\\
$^{126}$ Universit\"{a}t M\"{u}nster, Institut f\"{u}r Kernphysik, M\"{u}nster, Germany\\
$^{127}$ Universit\'{e} Clermont Auvergne, CNRS/IN2P3, LPC, Clermont-Ferrand, France\\
$^{128}$ Universit\'{e} de Lyon, CNRS/IN2P3, Institut de Physique des 2 Infinis de Lyon, Lyon, France\\
$^{129}$ Universit\'{e} de Strasbourg, CNRS, IPHC UMR 7178, F-67000 Strasbourg, France, Strasbourg, France\\
$^{130}$ Universit\'{e} Paris-Saclay, Centre d'Etudes de Saclay (CEA), IRFU, D\'{e}partment de Physique Nucl\'{e}aire (DPhN), Saclay, France\\
$^{131}$ Universit\'{e}  Paris-Saclay, CNRS/IN2P3, IJCLab, Orsay, France\\
$^{132}$ Universit\`{a} degli Studi di Foggia, Foggia, Italy\\
$^{133}$ Universit\`{a} del Piemonte Orientale, Vercelli, Italy\\
$^{134}$ Universit\`{a} di Brescia, Brescia, Italy\\
$^{135}$ Variable Energy Cyclotron Centre, Homi Bhabha National Institute, Kolkata, India\\
$^{136}$ Warsaw University of Technology, Warsaw, Poland\\
$^{137}$ Wayne State University, Detroit, Michigan, United States\\
$^{138}$ Yale University, New Haven, Connecticut, United States\\
$^{139}$ Yonsei University, Seoul, Republic of Korea\\
$^{140}$  Zentrum  f\"{u}r Technologie und Transfer (ZTT), Worms, Germany\\
$^{141}$ Affiliated with an institute covered by a cooperation agreement with CERN\\
$^{142}$ Affiliated with an international laboratory covered by a cooperation agreement with CERN.\\

\end{flushleft} 

\end{document}